\documentstyle[11pt,aaspp4]{article}
 
\newcommand{\co}{\rm CO}
\newcommand{\oo}{{\rm O}_2}
\newcommand{\hh}{{\rm H}_2}

\newcommand{\kps}{\,\textstyle\rm{km~s}^{-1}}
\newcommand{\Mpc}{\,\textstyle\rm{Mpc}}
\newcommand{\kpc}{\,\textstyle\rm{kpc}}
\newcommand{\coa}{CO(1$\rightarrow$0) }
\newcommand{\ooa}{O$_2(1,1\rightarrow 1,0)$ }
\newcommand{\yr}{\,\textstyle\rm{yr}}

\newcommand{\msun}{\,M_{\sun}}

\newcommand{\Jy}{\,\textstyle\rm{Jy}}
\newcommand{\K}{\,\textstyle\rm{K}}
 
\begin{document}

\title{MOLECULAR GAS IN MRK~109: CONSTRAINING THE O$_2$/CO RATIO IN
CHEMICALLY YOUNG GALAXIES}

\author{D. T. Frayer}\affil{Astronomy Department, California Institute
of Technology, Pasadena, CA  91125}
\author{E. R. Seaquist}\affil{Astronomy
  Department, University of Toronto, Toronto, ON, M5S 3H8, Canada}
\author{T. X. Thuan}\affil{Astronomy Department, University of Virginia,
Charlottesville, VA  22903}
\author{and} 
\author{A. Sievers}\affil{Institut de Radioastromie Milli\'{e}trique,
Granada, Spain}

\begin{abstract}

We report on observations of $^{12}$\coa emission from the chemically
young starburst galaxy Mrk~109.  These observations were part of a study
to constrain the O$_2$/CO ratio in metal--deficient galaxies, which were
motivated by theoretical work that suggests the possible enhancement of
the O$_2$/CO ratio in chemically young systems.  Five low metallicity
($Z \leq 0.5 Z_{\odot}$) IRAS galaxies at redshifts $z > 0.02$ (required
to shift the 118.75 GHz $^{16}$O$_2$ line away from the atmospheric
line) were searched for CO emission.  We detected the CO line in only
Mrk~109.  From O$_2$ observations of Mrk~109, we achieved an upper limit
for the O$_2$ column density abundance ratio of $N(\oo)/N(\co) < 0.31$.
These results provide useful constraints for the theoretical models of
chemically young galaxies.  We argue that either most of the molecular
gas in Mrk~109 does not reside in dark clouds ($A_{V} \ga 5$), or the
standard equilibrium chemistry models are inadequate for metal--poor
systems.  The molecular gas mass implied by the CO observations of
Mrk~109 is $M({\rm H}_2) \simeq 4 \times 10^{9} \msun$, and the CO data
are consistent with a central starburst induced by the interaction with
a nearby companion.

\end{abstract}

\keywords{galaxies: individual (Mrk109) --- galaxies: ISM --- galaxies:
  starburst}

\section{INTRODUCTION}

Although studying the molecular gas content of metal--deficient galaxies
is challenging, such efforts are essential for our understanding of the
formation and evolution of galaxies.  By studying chemically young
galaxies in the local universe, we gain insight into the processes which
occurred at early times ($z\sim 1-5$) for the metal--rich spiral and
elliptical galaxies found at the current epoch.  Unfortunately,
chemically young galaxies tend to be dwarf galaxies and are difficult to
detect in CO (Combes 1985; Arnault et al. 1988; Sage et al. 1992;
Israel, Tacconi, \& Baas 1995).  Due to the observational difficulties,
very little is currently understood about the molecular gas content of
chemically young galaxies.  One of the main uncertainties is the CO to
H$_2$ conversion factor.  Wilson (1995) and Arimoto, Sofue, \& Tsujimoto
(1996) have found an empirical relationship indicating an increase in
the CO to H$_2$ conversion factor with decreasing metallicity.  However,
detailed studies of the molecular clouds in the LMC and SMC suggest
other factors, besides metallicity, have an important role on the CO to
H$_2$ conversion factor (Rubio 1997; Israel 1997).

Although significant questions remain in our understanding of the CO to
H$_2$ factor, the molecular chemistry that occurs in metal--deficient
galaxies is even more uncertain.  Molecular chemistry calculations are
quite complicated and typically involve networks of hundreds to several
thousand reactions (Graedel, Langer, \& Frerking 1982; Herbst \& Leung
1989 [HL89]; Langer \& Graedel 1989 [LG89]; Bergin, Langer, \& Goldsmith
1995 [BLG95]).  Since the chemistry models still have difficulties in
explaining several of the observed abundance ratios in Galactic
molecular clouds, very little modeling has been devoted to metal--poor
galaxies.  One notable exception is the theoretical study of the
chemistry in LMC and SMC molecular clouds by Millar \& Herbst (1990)
[MH90].  In general, the molecular chemistry models are sensitive to a
variety of parameters, such as the density, temperature, and ionization
field, but the dominant parameter for determining the relative molecular
abundances of the carbon and oxygen species is the carbon to oxygen
ratio (LG89).  The models predict that the O$_2$/CO ratio decreases
exponentially with increases in the C/O ratio (LG89).  This could have
interesting ramifications on the molecular abundances in chemically
young galaxies.  Observations with the Hubble Space Telescope have shown
that the C/O abundance ratio increases with increasing metallicity in
{\sc Hii} galaxies (Garnett et al. 1995).  Therefore, we could expect to
find lower C/O ratios and correspondingly larger O$_2$/CO ratios within
dark molecular clouds in chemically young galaxies.  Based on these
simple ideas, the evolution of the O$_2$/CO abundance ratio as a
function of metallicity in galaxies has been recently quantified for a
variety of conditions and parameters governing the IMF and star
formation histories (Frayer \& Brown 1997 [FB97]).  At low
metallicities, FB97 calculate lower C/O ratios and enhanced O$_2$/CO
ratios (O$_2$/CO $\sim 1$) within dark ($A_{V} > 5$) molecular clouds.
At solar metallicities and above, the O$_2$/CO ratio is expected to
decrease by several orders of magnitude.

Molecular oxygen has yet to be detected conclusively outside our solar
system.  Due to atmospheric attenuation, the ground--based Galactic
searches have been limited to observing the rarer $^{16}$O$^{18}$O
isotope (Liszt \& Vanden Bout 1985 [LV85]; Goldsmith et al. 1985; Combes
et al.  1991 [C91]; Fuente et al. 1993; Mar\'{e}chal et al. 1997 [M97]).
The most sensitive Galactic studies have provided upper limits of
approximately O$_2$/CO $<0.1$.  Extragalactic searches for the
redshifted $^{16}$O$_2$ lines have provided more sensitive limits
(O$_2$/CO $<0.01$) but have also been unsuccessful (Liszt 1985,
Goldsmith \& Young 1989 [GY89], C91, Liszt 1992 [L92]; Combes \& Wiklind
1995).  The most sensitive limit to date is O$_2$/CO $<0.002$
($1\sigma$) from an absorption line study toward the radio source
B0218+357 (Combes, Wiklind, \& Nakai 1997).  All of these previous
searches for O$_2$ have concentrated on chemically rich systems, or
molecular clouds of unknown metallicity.  Since the theoretical models
suggest the possible enhancement of O$_2$ in chemically young galaxies,
we have carried out a search for O$_2$ in metal--deficient galaxies.

From the literature, we have compiled a list of metal--poor
($Z\leq0.5Z_{\sun}$) IRAS galaxies with redshifts of $z>0.02$.  There
are only about 20 galaxies satisfying these constraints, which were
primarily drawn from the samples of Salzer \& MacAlpine (1988) and
Dultzin--Hacyan, Masegosa, \& Moles (1990).  Unfortunately, most of
these galaxies are relatively weak IRAS sources $I(100\micron) \la
1$~Jy, and none have reported CO detections.  This is not terribly
surprising since metal--deficient galaxies have lower amounts of dust.
The metallicities for the galaxies in our sample were derived from their
oxygen abundances calculated using standard {\sc Hii} region analysis
techniques whenever the 4363\AA\,[{\sc Oiii}] line was observed
(Osterbrock 1989, case B).  For galaxies with no 4363\AA\,line, the
techniques of McGaugh (1991) were used to estimate the oxygen abundance.
In this initial study, we observed the strongest IRAS sources satisfying
the metallicity and redshift constraints.

\section{OBSERVATIONAL RESULTS}

\subsection{NRAO 12m Telescope}

In April 1996 we observed five of the strongest IRAS galaxies which are
known to be metal deficient and are at redshifts $z>0.02$ with the
NRAO\footnote{The National Radio Astronomy Observatory is a facility of
the National Sciences Foundation operated under cooperative agreement by
Associated Universities, Inc.} 12m telescope.  The redshift constraint
was required to permit the observations of the ground state transition
of $^{16}\oo$ N,J$=1,1 \rightarrow 1,0$ at a rest frequency of 118.7503
GHz.  For all five of the galaxies, we first searched for $^{12}$\coa
line emission.

With the 12m telescope system, we used dual polarization SIS mixer
receivers tuned to the frequency of the redshifted lines.  The spectra
were formed as total power differences between the source and a
reference position 4$\arcmin$ away in azimuth using a nutating
subreflector to switch at a rate of 1.25 Hz.  Each polarization was
recorded using a $256\times 2$ MHz filter bank spectrometer.  The
frequency setups were checked with observations of lines in Orion-KL and
Sag--B2 (Turner 1989).  Pointing was checked every 2--4 hours and was
found to be accurate to within $10\arcsec$.  Based on observations of
Mars, we derived a conversion factor of $40\pm 8$ Jy/K for the
$T_{R}^{*}$ temperature scale. This conversion factor assumes that the
source is smaller than the $60 \arcsec$ beam, which is valid for our
sample of galaxies.

Table~1 shows the \coa results of the 12m observations.  We
detected \coa emission only in Mrk~109.  The upper limits given in
Table~1 for the \coa line strengths of the other four galaxies are
$1\sigma$ rms, assuming a FWHM line width of $100\kps$.  For the
nondetections, the integration times range from 3 to 8 hours, and the
system temperatures were from 250~K to 300~K.  After the detection of
\coa emission in Mrk~109, we searched for the 118.75 GHz $^{16}$\ooa
line.  From these observations we marginally detected a feature
consistent with the O$_2$ line (Fig. 1).  This feature was seen in both
polarizations and in different subsets of the data.  Interestingly, at
the distance of Mrk~109 the \ooa line is redshifted to 115.22 GHz, which
is very near the \coa rest frequency.  However, the broad line--width
($\sim 200 \kps$) rules out Galactic clouds as a possible origin of the
115.22 GHz feature, and there are no known low--redshift galaxies within
or near the beam of the 12m observations.  The velocity agreement for
both features is encouraging (Fig. 1) and is consistent with both lines
being associated with Mrk~109.  Although 12 hours of source integration
time were invested for each of these lines, the \coa and \ooa lines
could only be considered tentatively detected with the 12m data.
Follow--up observations with more sensitive telescopes were required to
confirm the presence of these lines.

\subsection{IRAM 30m Telescope}

In July 1997 we obtained follow--up observations of Mrk~109 with the
IRAM\footnote{The Institut de RadioAstronomie Millim'{e}trique (IRAM)
is an international institute for research in millimeter astronomy,
cofunded by the Centre National de la Recherche Scientifique, France,
the Max Planck Gesellschaft, Germany, and the Instituto Geografico
Nacional, Spain.} 30m telescope.  With the 30m telescope system, we
simultaneously observed the redshifted \coa and \ooa lines.  Each line
was observed using a $512\times 1$ MHz filter bank detector.  The
pointing of the telescope was checked every 2 hours with observations of
0923+392 and was found to have an accuracy of better than $5\arcsec$.
The frequency setup of the observations was confirmed by test
observations of IRC+10216.  In three evenings we collected 6.3 hours of
integration time on Mrk~109.  Based on measurements of Mars taken each
day, we derived a conversion factor of $8.8\pm1.6$ Jy/K for the observed
$T_{A}^{*}$ temperature scale.

Figure~2 shows the spectra for the redshifted \coa and \ooa lines.
The \coa line was clearly detected, while the \ooa line was not.  Since
the optical size of Mrk~109 is less than the 30m beam size of
$21\arcsec$, we conclude that the 115.22 GHz feature apparent in the 12m
data does not arise from Mrk~109.  We cannot rule out the possibility
that this feature is associated with molecular gas outside the beam of
the 30m, but within the 12m beam.  One possibility, albeit improbable,
is that the 12m feature is due to \coa emission from a previously
undetected galaxy at $\sim 100 \kps$.  Most likely the 115.22 GHz
feature seen in the 12m data is spurious.

Based on the 30m data, we measure an integrated \coa line strength of
$7.9\pm 1.6 \Jy\kps$ over $cz = 9030-9230 \kps$ for Mrk~109, which is 10
times larger than the integrated rms noise across the profile.  The
uncertainty in this measurement is dominated by the absolute flux
calibration uncertainty of 18\%.  The $1\sigma$ upper limit to the \ooa
line strength is $I(\oo)/I(\co) < 0.15$.  We discuss the significance of
this upper limit in \S\ref{sec-o2co}.

\subsection{OVRO Imaging}

In January 1997, we obtained \coa and \ooa imaging observations of
Mrk~109 with the six element Owens Valley Radio Observatory (OVRO)
array.\footnote{The OVRO Millimeter Array is operated as a radio
astronomy facility by the California Institute of Technology.}  The \coa
line was observed in the lower side--band, while the \ooa line was
observed simultaneously in the upper side--band.  Each line was observed
using a $60\times 4$~MHz digital correlator spectrometer centered on the
redshifted frequencies of the lines.  The phase center of the
interferometric observations was the optical position of $\alpha$(B1950)
= $09^{\rm h} 19^{\rm m} 05\fs0$; $\delta$(B1950) =$+47\arcdeg 27\arcmin
28\arcsec$.  These observations were carried out in a low resolution
configuration (baselines ranging from 15--115 m), which provided a
resolution of $4\arcsec$ using natural weighting.  We obtained 20 hours
of effective integration time on Mrk~109 spread over 4 separate nights.
Observations of 3C~273 were used for passband and flux calibration.  The
observations of the nearby radio source 0923+392 were made every 20
minutes for gain calibration.  The data were calibrated using the OVRO
MMA software package, and the data analysis was accomplished using the
NRAO AIPS software package.

Figure~3 shows a natural--weighted map of the \coa emission in
Mrk~109 averaged over the velocity range showing CO emission in the 30m
data.  The \coa emission was unresolved and is spatially and
kinematically consistent with the optical regions in Mrk~109 (Mazzarella
\& Boroson 1993).  We measure an integrated \coa line strength of
$6.9\pm 1.8 \Jy\kps$ from the OVRO data which is consistent within
uncertainties with the single--dish measurements.  The line was detected
at the $5.7\sigma$ level in the OVRO data, and a major contributor to
the uncertainty of the CO line flux comes from the 20\% flux calibration
uncertainty.  Figure~4 shows the \coa and \ooa spectra taken along the
velocity axis of the data cubes at the peak position in the integrated
CO map.  The OVRO \coa line profile is very similar to that of the 30m.
We find no evidence of a O$_2$ line in Mrk~109 from the OVRO data.  No
other features, besides \coa emission at the position of Mrk~109, were
detected in the OVRO data cubes.

\section{DISCUSSION}

In this paper we report the detection of \coa emission in Mrk~109 as
well as the nondetection of CO in four distant metal--deficient galaxies
(Table~1).  We estimate the amount of molecular gas implied by these
observations using the metallicity dependent relationship derived
empirically by Wilson (1995).  Wilson finds that the CO to H$_2$
conversion factor increases with decreasing metallicity as a $-0.67$
power law.  We adopt this power law and calculate the molecular gas mass
(including He) using
\begin{equation}
M({\rm H}_2) = 1.6 \times 10^{4}\left(\frac{Z}{Z_{\sun}}\right)^{-0.67}
\left(\frac{d}{\Mpc}\right)^{2}S_{\rm CO}\,\msun,
\end{equation}
where S$_{\rm CO}$ is the CO flux in $\Jy \kps$(Wilson 1995).  This
relationship is consistent with simple theoretical arguments.  In the
low metallicity limit in which the CO emission becomes optically thin,
we would expect the conversion factor to increase roughly linearly with
decreasing metallicity (Sakamoto 1996).  As the metallicity reaches the
solar value, the CO emission becomes optically thick, and the conversion
factor becomes independent of metallicity, ignoring other physical
parameters.  For global observations of metal--deficient galaxies, we
could expect a mixture of CO optical depths and an effective power law
for the conversion factor between 0 and $-1$.

The S(CO)/S(100$\micron$) ratio for this sample of galaxies is
significantly smaller than that for normal elliptical and spiral
galaxies.  By using the CO to IRAS flux correlation derived for normal
galaxies (Bregman, Hogg, \& Roberts 1992), we would expect \coa fluxes
of $20-30 \Jy\kps$, which are much larger than those found for these
galaxies.  This is not a new observational result for metal--deficient
galaxies (e.g., Sage et al. 1992).  Metallicity by itself, however, does
not appear to explain this discrepancy.  Assuming an intrinsic gas to
dust ratio which increases linearly with decreasing metallicity (Issa,
MacLaren, \& Wolfendale 1990) and using the metallicity--corrected CO to
H$_2$ conversion factor (as in eq.[1]), we would expect the
S(CO)/S(100$\micron$) ratio to vary with $(Z/Z_{\sun})^{-0.33}$.  This
dependence is much weaker than that which is observed.  Clearly, more
work is needed before we can gain a complete understanding of the low
S(CO)/S(100$\micron$) ratios in metal--deficient galaxies.  For the
remaining of the paper, we focus on Mrk~109.

\subsection{Molecular Gas in Mrk~109}

The galaxy Mrk~109 is one of the more luminous metal--deficient
starburst galaxies.  It has an absolute magnitude of $M\sim -20$ which
places it apart from the more typical low--luminosity metal--poor dwarf
galaxies with $M\ga -18$ (French 1980).  Mrk~109 appears to be
interacting with a nearby companion located approximately $10\arcsec$
west of Mrk~109.  Both Mrk~109 and its companion are spectroscopically
classified as {\sc Hii} galaxies and are separated by only $100\kps$ in
radial velocity (Mazzarella \& Boroson 1993).

Figure~2 shows the velocities of the optical knots in Mrk~109 (labeled
a, b, c, \& d) and that of the companion galaxy (labeled A).  The CO
emission spans the full range of velocities given by the optical
components, which is consistent with the interaction between Mrk~109 and
the companion galaxy.  Interestingly, the \coa profile peaks at the
velocity of the companion galaxy, while the CO emission is spatially
consistent with the brightest central optical knot (b) in Mrk~109.  The
interpretation of these results is not clear.  It is possible that the
kinematic center of Mrk~109 is offset from the brightest optical region
and is more consistent with the average velocity of knots (a) and (d).
Alternatively, the bulk of the gas observed may have been stripped from
the companion galaxy (A).  Unfortunately, we lack the sensitivity and
resolution to provide a detailed kinematic study of the molecular gas in
this system.  However, we can deduce general characteristics such as the
star--formation efficiency (SFE) and the gas fraction for this
merger/starburst system.

Based on the 30m data, we derive a molecular gas mass of $M({\rm H}_2) =
(3.8\pm0.7) \times 10^{9} h_{75}^{-2} \msun$ for Mrk~109, using
equation~(1) where $h_{75} = {\rm H}_o/(75 \kps \Mpc^{-1})$.  This is a
relatively large amount of molecular gas for a metal--deficient galaxy
and is similar to the total amount of molecular gas estimated for the
Milky Way ($2\times 10^{9}\msun$, Solomon \& Rivolo 1987).  The
dynamical mass contained within the CO emission regions is $M_{dyn}=
R(\Delta V/[2\sin(i)])^2/G $, where $\Delta V$ is the
observed FWZI line width of $200 \kps$. The resolution of the OVRO data
places an upper limit on the size of the CO emission region of $R < 1.2
h_{75}^{-1} \kpc$.  The inclination of Mrk~109 is unknown.  Assuming an
intrinsic axial ratio of one, the observed optical axial ratio
(Mazzarella \& Boroson 1993) implies an inclination of $i\sim
61\arcdeg$.  Using the above numbers, we calculate a dynamical mass of
$M_{dyn} < 3.6 \times 10^{9} h_{75}^{-1} \msun$.  These results suggest
a gas fraction of approximately unity within the central regions of
Mrk~109.  Granted, given the uncertainties in the CO to H$_2$ conversion
factor and inclination, the gas fraction may be lower.  If we,
nevertheless, take these results at face value, the large calculated gas
fraction is consistent with a young starburst, where the majority of the
gas has yet to be consumed by the ongoing star formation.

By using the H$\alpha$ line luminosity of Mrk~109 (Mazzarella, Bothun,
\& Boroson 1991), we estimate a total star--formation rate (SFR) of $2
\msun\yr^{-1}$ (Kennicutt 1983).  This value is consistent with the SFR
expected from its far infrared luminosity, assuming the relationship
given by Sage et al. (1992).  In addition, the derived SFR is also
consistent with the upper limit implied from the radio data.  Mrk~109
has not been detected at cm wavelengths.  The data from the NRAO VLA Sky
Survey (Condon et al. 1998) implies a flux density of less than 1~mJy
($2\sigma$) at 1.4~GHz.  Using this value we derive an upper limit to
the total SFR of approximately $2\msun\yr^{-1}$, assuming non--thermal
radio emission from supernovae remnants and renormalizing the formulae
of Condon (1992) to match the IMF of Kennicutt (1983).  The consistency
of the radio, infrared, and optical data indicates that the SFR has not
been seriously underestimated due to high--extinction regions that are
hidden at optical wavelengths.

The star--formation efficiency can be defined as the the ratio of the
SFR to the amount of molecular gas.  For Mrk~109, we find an implied
efficiency of only SFE$\sim 0.5$ Gyr$^{-1}$, which is an order of
magnitude lower than that typically found for starburst galaxies (Sage
et al. 1992).  Assuming a constant SFR, this would translate into a gas
depletion time scale of 2~Gyr for Mrk~109.  The SFR, SFE, and gas
depletion time scale of Mrk~109 are, hence, more similar to those found
for normal spiral galaxies than starburst systems.  At first glance,
these results could appear contradictory with a merger scenario where we
expect to typically find high SFRs (e.g., Mihos, Richstone, \& Bothun
1992).  However, most previously studied merger systems are older
metal--rich, IRAS luminous starbursts (e.g., Sanders, Scoville, \&
Soifer 1991).  Mrk~109 is a low metallicity system whose ISM has
undergone significantly less processing.  It is possible that Mrk~109 is
a young interacting system still in the process of gaseous infall into
the central starburst region which could explain its current low SFE.

\subsection{Constraining the O$_2$/CO Ratio for Sub--solar
Metallicities}
\label{sec-o2co}

The major goal of this study is to constrain the O$_2$/CO abundance
ratio for chemically young systems.  Previous searches for O$_2$
emission have been directed toward CO bright sources, such as nearby
giant molecular cloud (GMC) cores or the centers of bright IRAS galaxies
(\S1).  In this paper, we present the first O$_2$/CO abundance limit for
a metal--deficient galaxy.

The observed intensity limit for the \ooa line in Mrk~109 is
$I_{\oo}/I_{\co} < 0.15$ (\S2.2).  By combining equation~(1) with the
expression given by Liszt (1992), we can relate the observed intensity
ratio to a column density abundance ratio with the following expression
(FB97):
\begin{equation}
\frac{I(\oo)}{I(\co)} =0.68 \left(\frac{N(\oo)}{N(\co)}\right)
\left(\frac{Z}{Z_{\sun}}\right)^{0.33} 
\left(\frac{30\K}{T}\right).
\end{equation}
The $N(\oo)/N(\co)$ ratio is the column density abundance ratio averaged
over the telescope beam.  It is related to the intrinsic O$_2$/CO
abundance ratio by $N(\oo)/N(\co) = (\oo/\co)(f_{\oo}/f_{\co})$, where
$f_{\oo}/f_{\co}$ is the ratio of the filling factors of the O$_2$ and
CO emission regions within the telescope beam.  If the CO and O$_2$
emission regions are coextensive, then $f_{\oo}/f_{\co}=1$.  Although
this has been assumed for all previous O$_2$ studies, it may not be
valid.  For now we adopt $f_{\oo}/f_{\co}=1$ and discuss the
applicability of this assumption later.

By adopting $T=30$~K and using equation~(2), we calculate $N(\oo)/N(\co)
< 0.31$ ($1\sigma$) for Mrk~109.  To determine the significance of this
upper limit, we have compiled a list of observational limits in regions
of known metallicities and/or O/C abundance ratios.  Figure~5 compares
the observational upper limits of the O$_2$/CO ratio with theoretical
expectations as a function of the O/C ratio.  We first summarize the
data displayed in Figure~5, and then we discuss several implications of
the data.

The analytical fits in Figure~5 were made to the steady--state and
early--time solutions of LG89 and MH90 (FB97).  The steady--state
solutions represent the equilibrium values achieved in the chemistry
models after $t>10^{7}\yr$, while the early--time solutions are for
$t\sim 10^{5} \yr$.  The LG89 solutions are for Galactic clouds, and the
MH90 solutions are for clouds in the LMC and SMC.  Both sets of
calculations assume dark clouds ($A_{V} \sim 10$).  Figure~5 shows
the effects of different processes on the steady--state solutions, such
as grain surface adsorption and desorption (BLG95), turbulent diffusion
(Xie, Allen, \& Langer 1995 [XAL95]), and mixing (Pineau des For\^{e}ts,
Flower, \& Chi\`{e}ze 1992 [PFC92]).

Most of the previous searches for O$_2$ have been made for regions of
unknown O/C abundance ratios and uncertain metallicities, and the
previous comparisons between the observational limits of the O$_2$/CO
ratio and the theoretical calculations generally assume a solar O/C
abundance ratio.  Since the O$_2$/CO ratio is strongly nonlinear with
the O/C ratio near solar metallicities, accurate estimates of the O/C
ratio are required in order to constrain the theoretical models.  We
find only three sources in the literature with both O$_2$/CO upper
limits and a direct estimate of the O/C ratio (Orion, L134N, and NGC
7674).  All three of these sources have sufficiently low O/C ratios that
their observational O$_2$/CO limits do not contradict the theoretical
estimates (Fig.~5).  In Figure~5 we adopt the average nebular abundance
ratio of O/C$=1.4$ for Orion (Cunha \& Lambert 1994 [CL94]).  For the
molecular complex L134N, we display the data at the position of the peak
in the SO map ($-1\farcm 4, -0\farcm 8$) which has the highest inferred
O/C ratio in L134N (O/C$=0.83$, Swade 1989 [S89]).  The O/C ratio for
the Seyfert galaxy NGC~7674 was based on the best model fit to its IUE
spectra (Kraemer et al. 1994 [K94]). For the data point marked with the
symbol ``$\ast$'' in Figure~5, we assume that the Galactic molecular
regions searched for O$_2$ (LV85; M97) have O/C ratios in agreement with
typical values observed in the diffuse ISM of O/C$\simeq 1.25 - 2.5$
(Cardelli et al. 1993 [C93]).  The assumption of similar O/C ratios in
the diffuse ISM and in regions of high extinction (e.g., cores of GMCs)
may not be valid, especially considering the results for L134N where
O/C$< 1$.  For the galaxies NGC~6090 and Mrk~109, we estimate the O/C
ratio assuming the relationship seen between the metallicity and the C/O
ratio (FB97; Garnett et al. 1995).  The metallicity for NGC~6090 was
derived by Storchi--Bergmann, Calzetti, \& Kinney (1994) [S94] using
indirect methods consistent with those of McGaugh (1991) and is fairly
uncertain (0.15 dex error).  The metallicity for Mrk~109 is more
reliable (Table~1) since the 4363\AA\,[{\sc Oiii}] line was observed
(French 1980).  The uncertainties in the O/C ratio derived for Mrk~109
and NGC~6090 reflect the scatter in the data for the relationship
between metallicity and the C/O ratio (see Fig.~7 of FB97).

All $^{16}\oo/\co$ abundance limits shown in Figure~5 are $2\sigma$.
For Galactic sources, the data are based on observations of the
$^{16}{\rm O}^{18}{\rm O}$ molecule and standard isotopic ratios are
assumed.  Estimated gas temperatures are used for the individual
Galactic sources (e.g., $T=10 - 40 \K$, M97) to account for the
temperature dependency on the O$_2$ abundance.  For the extragalactic
sources, we assumed $T= 30 \K$.  We used equation~(2) to derive the
O$_2$/CO abundance ratio for the galaxies Mrk~109 and NGC~6090.  For
NGC~7674 we used the Galactic CO to H$_2$ conversion factor; i.e., we
removed the $(Z/Z_{\sun})^{0.33}$ term from equation~(2) which is
applicable for only metal--deficient galaxies.  It was assumed that the
filling factors of the O$_2$ and CO emission regions are similar
($f_{\oo}/f_{\co} \sim 1$) for all of the limits in Figure~5.

By including the data for Mrk~109, we can begin to discuss several
implications for the O$_2$/CO abundance ratio over a relatively wide
range of O/C ratios (Fig.~5).  If we assume that Galactic molecular
clouds have a solar abundance ratio of O/C$\simeq 2$ (Grevesse, Noels,
\& Sauval 1996), the observations contradict the standard steady--state
solutions for molecular chemistry.  Several authors have already
summarized various possible mechanisms which could account for the
discrepancy (see FB97; M97).  One promising mechanism is mixing.  When
the dense interior gas mixes with the outer layers, the clouds do not
reach their steady--state chemistry solutions, but have solutions
similar to those derived at early times.  For these models the O$_2$/CO
ratios are expected to be lowered by several orders of magnitude at
solar abundances due to a higher fraction of the carbon in atomic and
ionized species (FB97).  For metal--deficient systems these effects
could be less important since the O/C ratios are larger.  In fact, for
O/C$ \ga 10$ the early--time solutions begin to converge with those of
steady--state.  In this scenario the metal--deficient galaxies could be
the best sources to search for O$_2$.

The observations of Mrk~109 provide us with the first probe of a
metal--deficient system.  The O$_2$/CO upper limit for Mrk~109 is lower
than the expected steady--state solutions (Fig.~5).  As stated earlier,
the theoretical solutions were derived assuming dark molecular clouds.
Since the photodissociation rate for O$_2$ is larger than that for CO
(van Dishoeck 1988), the global averaged O$_2$/CO abundance ratios could
be significantly lower than the dark cloud solutions (FB97).  Therefore,
at low metallicities we have two competing tendencies affecting the
global O$_2$/CO ratio.  On one hand, the production of O$_2$ is enhanced
due to the higher O/C ratios, but on the other hand the O$_2$ molecules
are more susceptible to photodissociation.  The effect that
photodissociation has on the global O$_2$/CO abundance ratio depends on
the fraction of the molecular gas residing in dark clouds.  Significant
O$_2$ emission is only expected from regions with $A_{V}\ga5$, while CO
emission can arise from regions with lower values of extinction (FB97).
Since the CO emission is thought to be more pervasive than the O$_2$
emission, we would expect relative filling factors of $f_{\oo}/f_{\co} <
1$ for extragalactic observations of a collection of clouds.  Hence, the
standard assumption of $f_{\oo}/f_{\co} =1$ is not valid in general.
Assuming spherical molecular clouds, we estimate that
$M(\hh)_{dark}/M(\hh)_{total} \simeq (f_{\oo}/f_{\co})^{1.5}$ based on
the work of FB97.  If we use the steady--state theoretical solutions, we
would expect O$_2$/CO $=1.3$ for the dark cloud regions ($A_{V} >5$) in
Mrk~109.  The $1\sigma$ upper limit on the $N(\oo)/N(\co)$ ratio for
Mrk~109 would then suggest $f_{\oo}/f_{\co} < 0.24$ and
$M(\hh)_{dark}/M(\hh)_{total} < 0.12$.  In summary, either most of the
molecular gas does not reside in dark clouds in Mrk~109, or the
steady--state solutions do not apply for Mrk~109.

By using the models of FB97 which assume the size--density relationship
and the GMC mass function found for Galactic clouds, we estimate
$f_{\oo}/f_{\co} \simeq 0.05$ for regions with a metallicity similar to
that of Mrk~109.  This value is calculated from the ratio of the
volume--weighted mass fractions of molecular hydrogen traced by the
O$_2$ and CO molecules; $f_{\oo}/f_{\co} =
(M[\hh]^{\oo}/M[\hh]^{\co})^{2/3}$ (FB97).  Arguably, molecular clouds
at low metallicities may be comprised of higher density clumps than
those found for Galactic GMCs, given the studies of the LMC and SMC
(Lequeux et al. 1994).  In such a case, the above value for the
$f_{\oo}/f_{\co}$ ratio should be thought of as a lower limit.  The
likely value of $f_{\oo}/f_{\co}$ for Mrk~109 is, therefore, $0.05 <
f_{\oo}/f_{\co} < 0.24$.  In this context, the observational O$_2$/CO
limit for Mrk~109 is not that restrictive on the models of molecular
gas--phase chemistry in dark clouds.  Both the steady--state and
early--time solutions are possible for values of $f_{\oo}/f_{\co} \sim
0.1$.

\section{CONCLUSIONS}

In this paper, we report the nondetection of \coa emission in four
distant metal--deficient galaxies and the detection of CO emission in
Mrk~109.  The \coa observations of Mrk~109 imply the presence of $4
\times 10^{9} \msun$ of molecular gas within the central few kpc of
Mrk~109.  The observational data favor a recent interaction of Mrk~109
with a nearby companion which has induced a starburst in the central
regions of Mrk~109.  The low metallicity, high gas fraction, and the
kinematics are consistent with a young starburst.

Based on the competing photodissociation and chemistry effects, we
expect the largest global O$_2$/CO ratios for systems with metallicities
in the range of $0.1 < Z/Z_{\sun} < 0.5$ (FB97).  Although the
metallicity of Mrk~109 falls within this optimal range, we fail to
detect O$_2$ emission in Mrk~109.  The observed intensity upper limit is
$I(\oo)/I(\co) < 0.15$.  These observations provide the first constraint
on the abundance of O$_2$ in a metal--deficient galaxy.  We derive a
column density abundance limit of $N(\oo)/N(\co)<0.31$ for Mrk~109.
These results suggest that most of the molecular gas in Mrk~109 does not
reside in dark clouds, and/or that the gas--phase steady--state
chemistry models do not apply for Mrk~109.  It would be challenging to
significantly improve the O$_2$/CO limit for chemically young
environments derived from this work with current ground--based
instrumentation.  Future observations with satellites, such as ODIN
(Hjalmarson 1997), of the LMC and SMC should provide more useful limits.

\acknowledgements

We are grateful to the staff of the NRAO 12m telescope, IRAM 30m
telescope, and OVRO mm--array who made these observations possible.  We
thank the referee C. Wilson for constructive suggestions concerning the
manuscript.  This research has made use of the NASA/IPAC Extragalactic
Database (NED) which is operated by the Jet Propulsion Laboratory,
California Institute of Technology, under contract with the National
Aeronautics and Space Administration.  This research was supported in
part by a grant to E.R.S. from the Natural Sciences and Engineering
Research Council of Canada and by the National Science Foundation grant
AST--96--13717 to the Owens Valley Radio Observatory.

\newpage
\figcaption[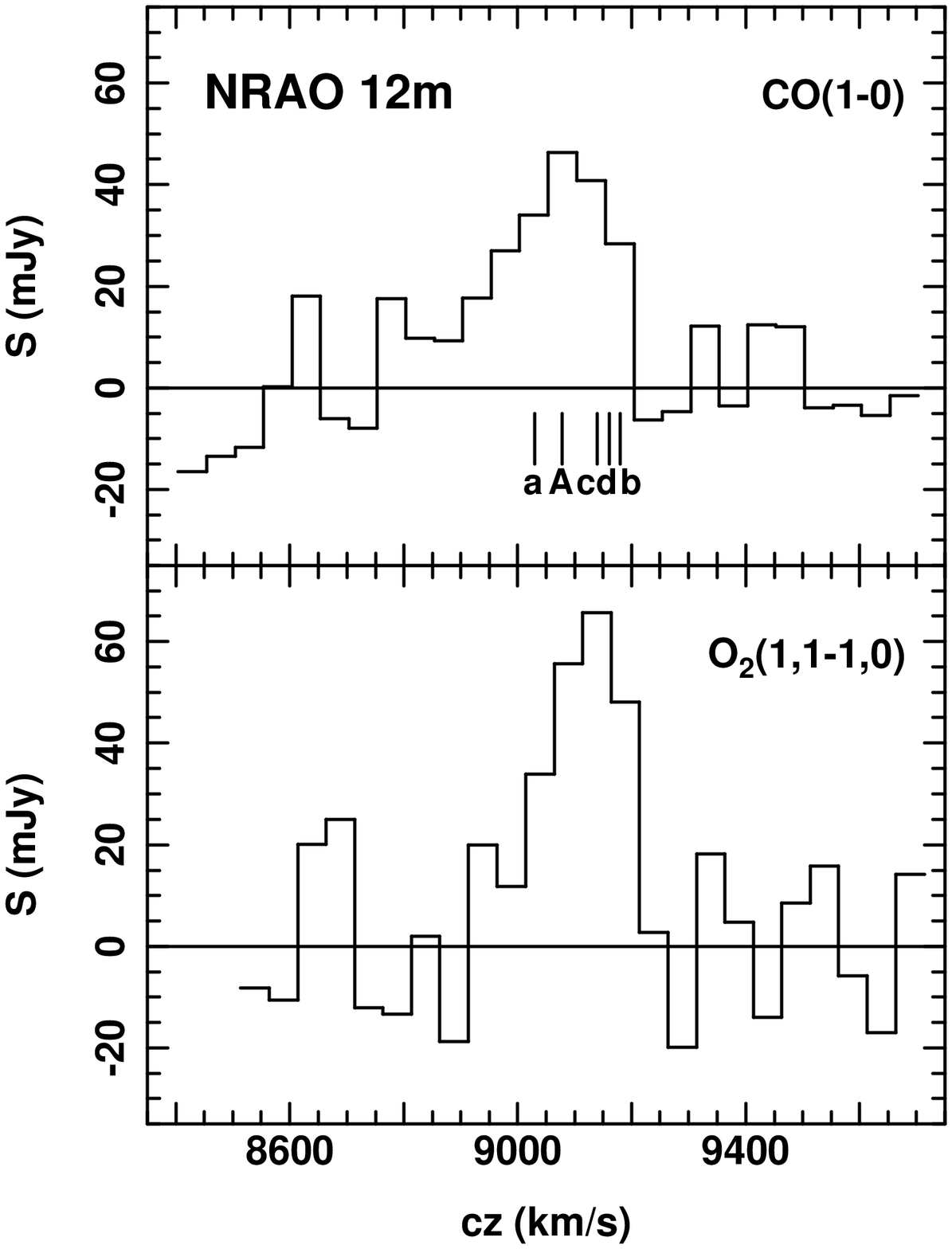]{The \coa and \ooa spectra of Mrk~109 observed with
the NRAO 12m telescope.  The heliocentric velocities of the optical
knots in Mrk~109 are labeled with ``a,b,c, \& d'', while the nearby
companion galaxy has a velocity labeled with ``A'' (Mazzarella \&
Boroson 1993).  The data have been smoothed to $50\kps$, and the
$1\sigma$ rms is 19 mJy and 25 mJy for the CO and O$_2$ data,
respectively.  The apparent \ooa feature is most likely spurious, given
the results shown in Fig.~2\&4.}

\figcaption[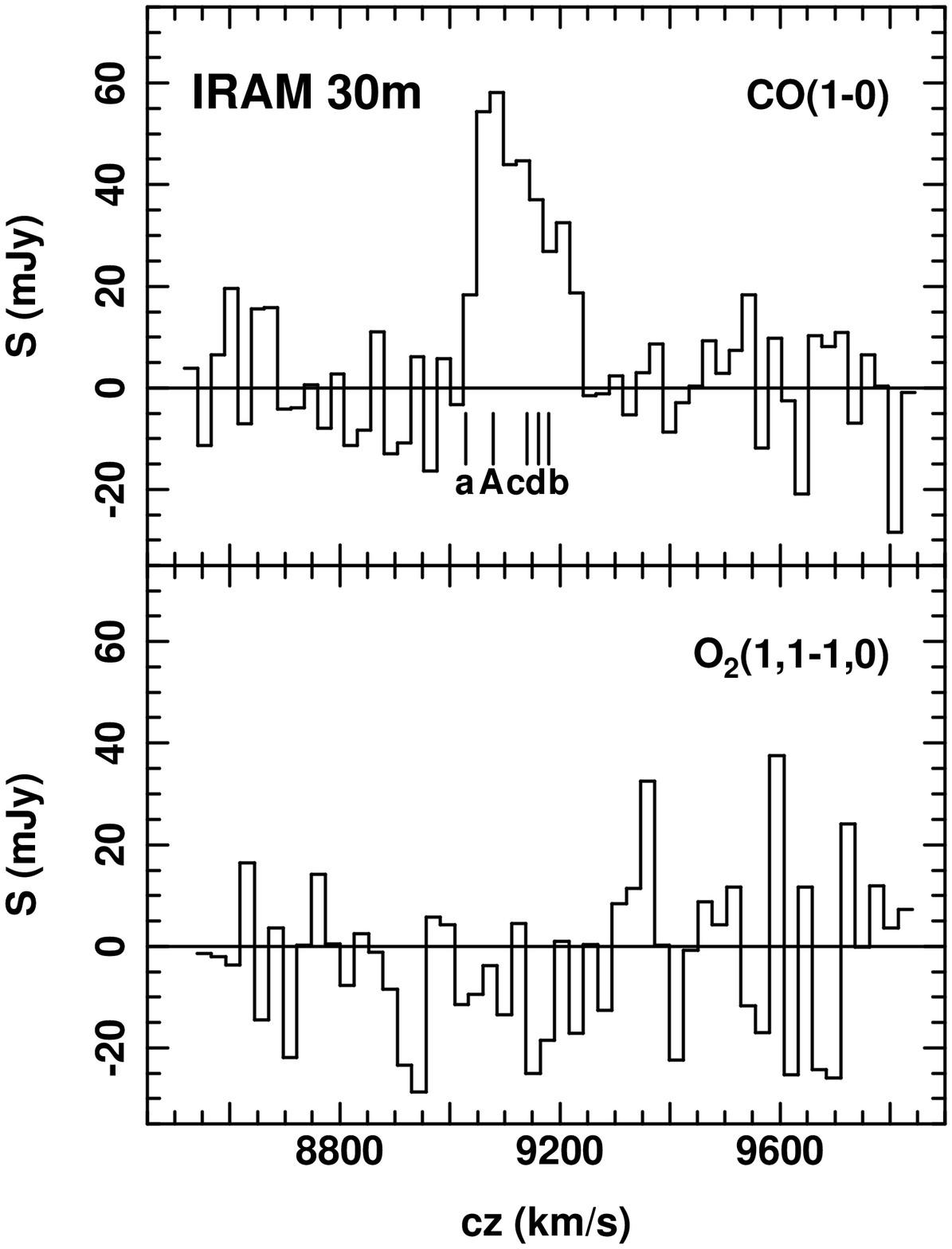]{The \coa and \ooa spectra of Mrk~109 observed with
the IRAM 30m telescope.  The labels for the velocities of optical
regions are the same as in Figure~1.  The data have been smoothed to
$25\kps$, and the $1\sigma$ rms is 11 mJy and 16 mJy for the CO and O$_2$
data, respectively.  No \ooa emission is detected.}

\figcaption[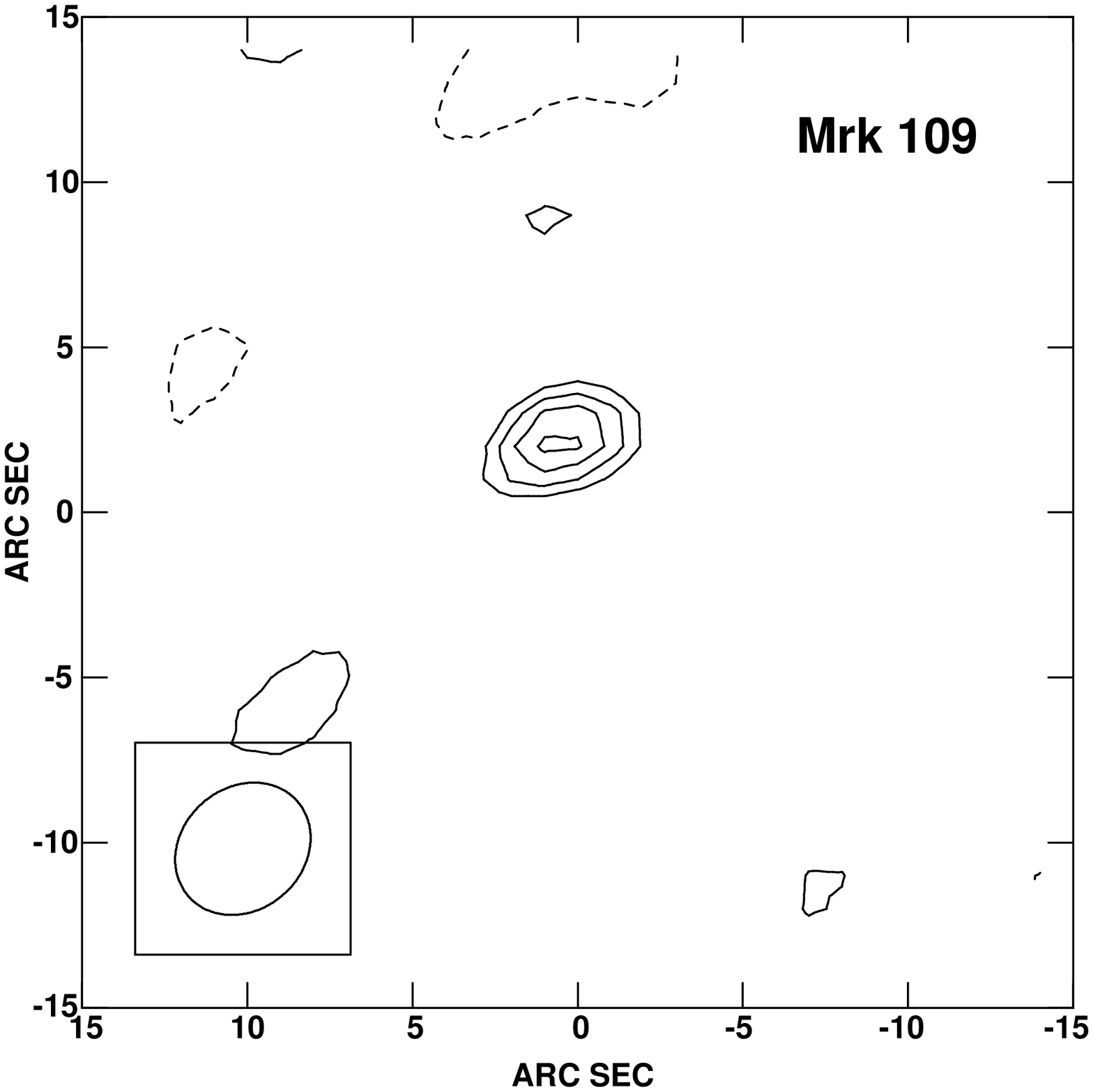]{The integrated \coa map of Mrk~109 observed with the
OVRO mm--array.  The CO emission was integrated over the velocity range
showing CO emission in the 30m spectrum (Fig.~2).  The positional
offsets are relative to the optical position of Mrk~109, and the
location of the CO emission is consistent with the brightest optical
regions in Mrk~109 (Mazzarella \& Boroson 1993).  The $1\sigma$ rms is
$1.2 \Jy \kps$/beam, and the contour levels are $-2.5\sigma$,
$2.5\sigma$, $3.5\sigma$, $4.5\sigma$, \& $5.5\sigma$.  The synthesized
beam size is shown at the lower left.}

\figcaption[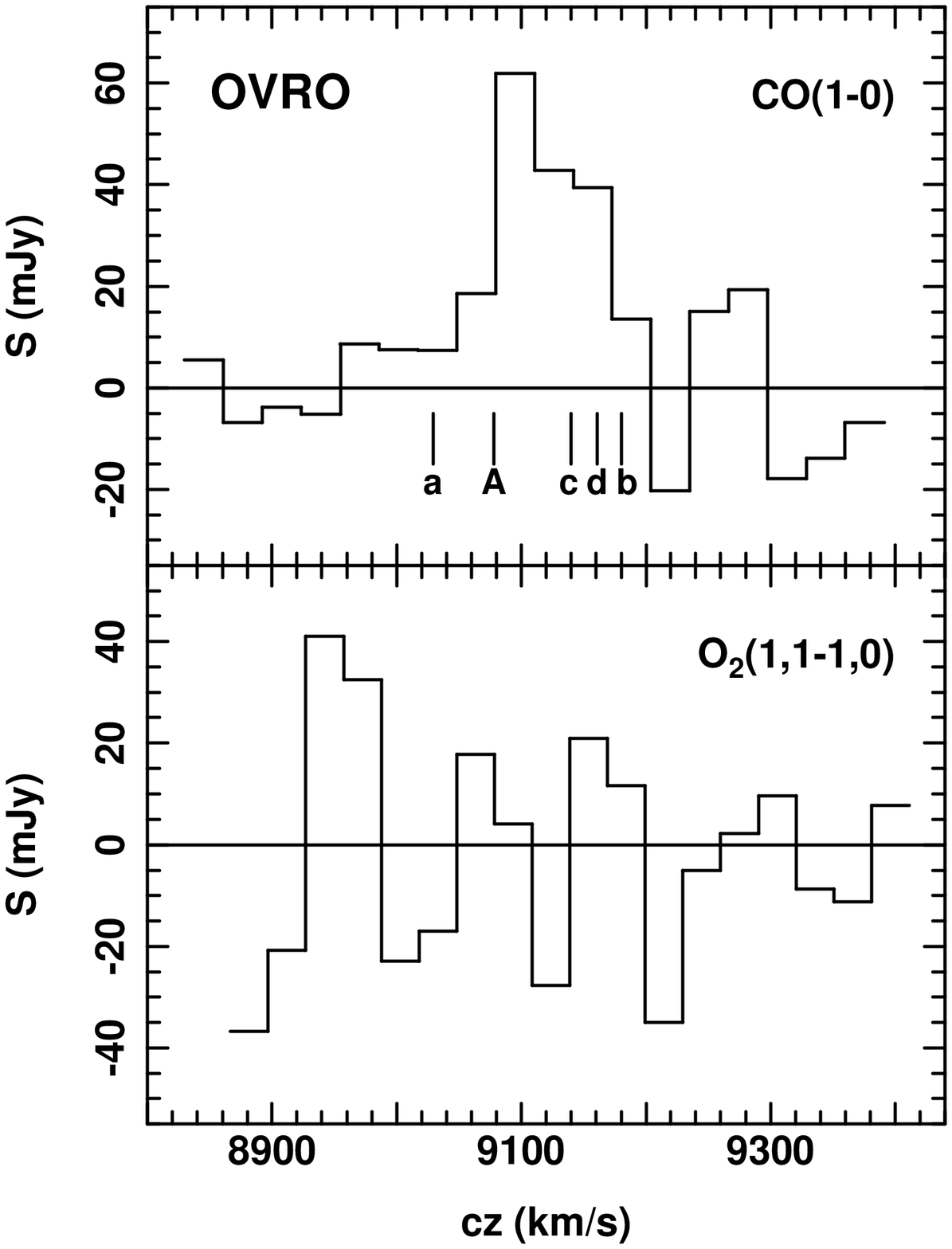]{The \coa and \ooa spectra of Mrk~109 at the peak
position in the integrated CO map (Fig. 3).  The labels for the
velocities of optical regions are the same as in Figure~1.  The data have
been smoothed to 12 MHz ($\sim 30 \kps$), and the $1\sigma$ rms is 14
mJy/beam and 19 mJy/beam for the CO and O$_2$ data, respectively.  No
\ooa emission is detected.}

\figcaption[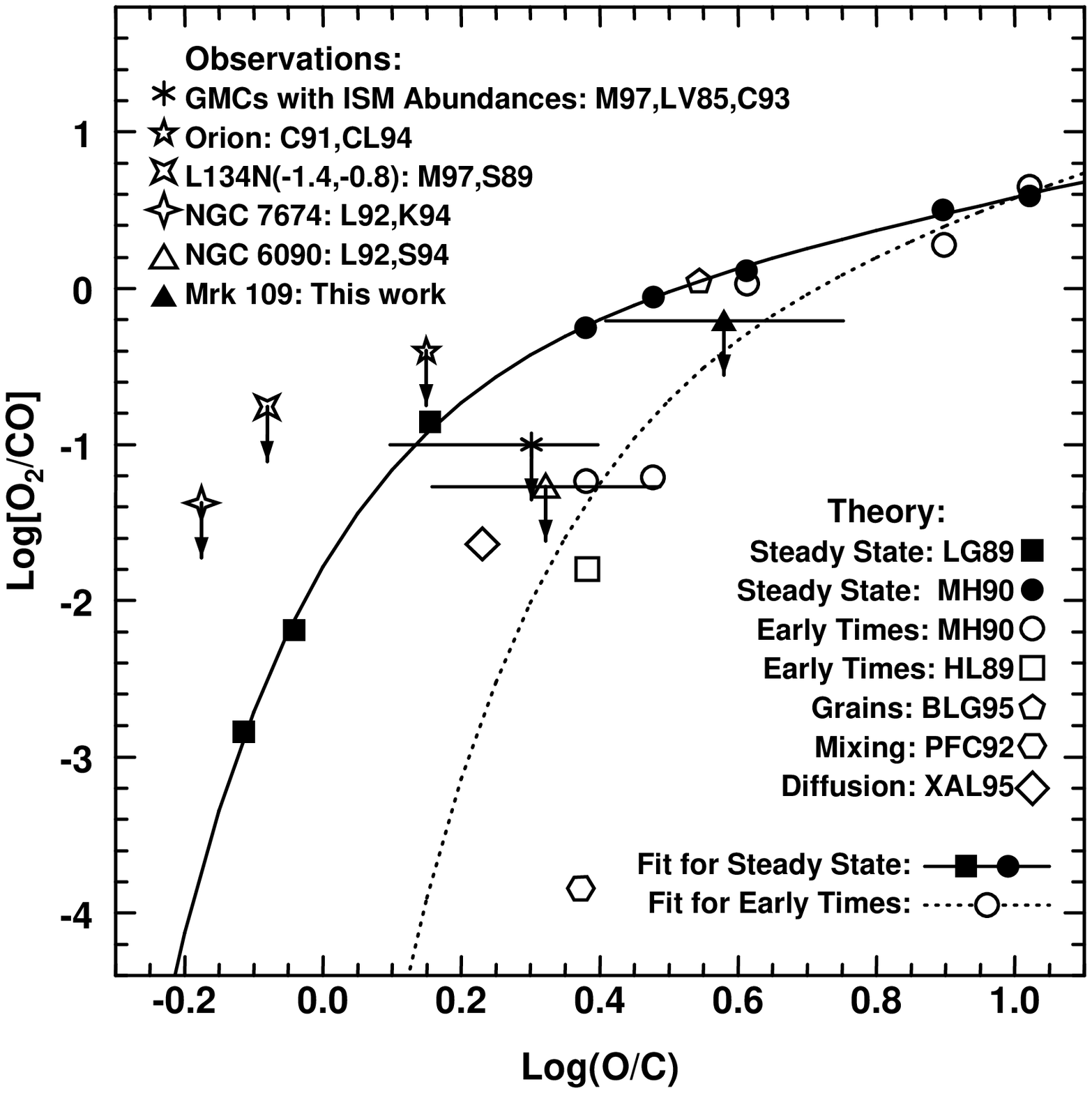]{The O$_2$/CO ratio as a function of the O/C ratio.
    Observational upper limits of the O$_2$/CO ratio ($2\sigma$) are
    shown along with theoretical data from several different time
    dependent molecular chemistry models (see text).  Analytical fits
    are shown for the steady--state solutions of LG89 and MH90 and for
    the early--time solutions of MH90.  The data point for Mrk~109 is
    the first observational constraint for the O$_2$/CO ratio in
    low--metallicity molecular gas.}

\newpage

\begin{deluxetable}{lcccrr}

\tablenum{1} \tablewidth{0pt} \tablecaption{\coa Observations at the 12m
Telescope} \tablehead{ \colhead{Name} & \colhead{Redshift}&
\colhead{Metallicity$^{\rm a}$}& \colhead{S(100$\micron$)} &
\colhead{\coa}& \colhead{M(H$_2)^{\rm b}$}\nl \colhead{ } &
\colhead{(z)}& \colhead{(O/H)/(O/H)$_{\sun}$}& \colhead{(Jy)}&
\colhead{($\Jy\kps$)}& \colhead{($10^{9}\,{\rm h}_{75}^{-2} \msun$)} }
\startdata

UM 17 & 0.0274 & 0.31$^{\rm c}$ & 1.6$^{\rm d}$ & $<2.4^{\rm e}$&$<1.0$
\nl 

Haro 15 & 0.0216 & 0.50$^{\rm f}$ & 2.0$^{\rm g}$& $<1.5^{\rm
e}$&$<0.3$\nl

Mrk 109 & 0.0306 & 0.36$^{\rm h}$ & 1.1$^{\rm g}$& $9.2 \pm 3.2^{\rm
i}$&$4.3\pm1.5$ \nl

Mrk 162 & 0.0215 & 0.11$^{\rm h}$ & 1.6$^{\rm g}$& $<2.1^{\rm e}$&
$<1.1$ \nl

UM 603 & 0.0300 & 0.42$^{\rm j}$ & 1.5$^{\rm k}$ & $<2.2^{\rm
e}$&$<0.9$ \nl \enddata

\tablecomments{$^{\rm a}$Based on a solar oxygen abundance of
$12+\log({\rm O/H}) = 8.87$ (Grevesse et al. 1996). $^{\rm b}$Molecular
gas mass calculated assuming the metallicity dependent CO to H$_2$
conversion factor of Wilson (1995); h$_{75} = {\rm H}_{o}/(75 \kps
\Mpc^{-1})$.  $^{\rm c}$Metallicity estimated from the data of Terlevich
et al. (1991) using the method of McGaugh (1991).  $^{\rm
d}$Dultzin--Hacyan et al. (1990). $^{\rm e}1\sigma$ rms assuming
$100\kps$ line width. $^{\rm f}$Storchi--Bergmann et al. (1994).  $^{\rm
g}$NASA/IPAC Extragalactic Data Base.  $^{\rm h}$Kunth \& Sevre
(1985). $^{\rm i}$Measured over 8880--9220 $\kps$ with the uncertainty
calculated from a combination of the 28\% rms uncertainty and the 20\%
absolute flux calibration uncertainty. $^{\rm j}$Campos--Aguilar, Moles, \&
Masegosa (1993). $^{\rm k}$Salzer \& MacAlpine (1990).}

\end{deluxetable}


\setcounter{figure}{0}

\newpage
\begin{figure}
\includegraphics{f1.ps} \vspace*{8.in}
\caption{ }
\end{figure}

\newpage
\begin{figure}
\includegraphics{f2.ps} \vspace*{8.in}
\caption{ }
\end{figure}

\newpage
\begin{figure}
\includegraphics{f3.ps} \vspace*{8.in}
\caption{ }
\end{figure}

\newpage
\begin{figure}
\includegraphics{f4.ps} \vspace*{8.in}
\caption{ }
\end{figure}

\newpage
\begin{figure}
\includegraphics{f5.ps} \vspace*{8.in}
\caption{ }
\end{figure}

\end{document}